\documentstyle[preprint,aps]{revtex}

\tightenlines

\begin{document}

\draft

\title{
Determination of the Kobayashi-Maskawa-Cabibbo matrix element $V_{us}$ under
various flavor-symmetry breaking models in hyperon semileptonic decays
}

\author{
R. Flores-Mendieta
\ \ \ and \ \ \
A. Garc\'\i a
}
\address{
Departamento de F\'\i sica, Centro de Investigaci\'on y de Estudios
Avanzados del IPN,\\
Apartado Postal 14-740, 07000, M\'exico, Distrito Federal, Mexico
}
\author{
G. S\'anchez-Col\'on
}
\address{
Departamento de F\'\i sica Aplicada, Centro de Investigaci\'on y de
Estudios Avanzados del IPN,\\
Unidad M\'erida, \\
Apartado Postal 73, Cordemex 97310, M\'erida, Yucat\'an, Mexico.
}

\date{February 20, 1996}

\maketitle

\begin{abstract}
\noindent We study the success to describe hyperon semileptonic decays of 
four models that incorporate second-order $SU(3)$ symmetry breaking 
corrections. The criteria to assess their success is by determining 
$V_{us}$ in each of the three relevant hyperon semileptonic decays and 
comparing the values obtained with one another and also with the one that 
comes from $K_{l3}$ decays. A strong dependence on the particular 
symmetry breaking model is observed. Values of $V_{us}$ which do 
not agree with the one of $K_{l3}$ are generally obtained. However, in the 
context of chiral perturbation theory, only the model whose corrections 
are $O(m_s)$ and $O(m_s^{3/2})$ is successful. Using its predictions for 
the $f_1$ form factors one can quote a value of $V_{us}$ from this 
model, namely, $V_{us}=0.2176\pm 0.0026$, which is in excellent 
agreement with the $K_{l3}$ one. 
\end{abstract}


\section{INTRODUCTION}

From the theoretical point of view, hyperon semileptonic decays (HSD) are
considerably more complicated than pseudoscalar-meson semileptonic decays.
The participation of vector and axial-vector currents in the former leads to
the appearance of many more form factors. While in the latter not only 
less form factors appear, because only the vector current can participate,
but also the theoretical approach to compute such form factors is under 
quite reasonable control\cite{leut}. These facts allow that the 
Kobayashi-Maskawa-Cabibbo matrix element $V_{us}$ be more reliably  
determined in the decays $K^+ \rightarrow \pi^0 l^+ \nu_l$ and $K^0 
\rightarrow \pi^- l^+ \nu_l$ than in HSD. An analysis of $K_{l3}$ 
decays\cite{leut} yields $V_{us} = 0.2196 \pm 0.0023$. The inclusion of 
more refined $SU(2)$ symmetry-breaking corrections leads to\cite{gabriel}

\begin{equation}
V_{us} = 0.2188 \pm 0.0016 \, .
\label{ec1}
\end{equation}

It is difficult to assess the success of the many calculations of $SU(3)$
symmetry-breaking corrections to the form factors of HSD. Predictions 
that vary substantially from one another are obtained. An important 
selection of such calculations can be found in 
references\cite{sch,krause,dono,ander}.
These are refined calculations that incorporated second-order symmetry
breaking corrections to the leading vector form factor $f_1$. However, a 
reliable knowledge of $V_{us}$ provides an opportunity to establish
some criteria to discriminate between the several such calculations.
If one uses them to determine $V_{us}$ from HSD, then --in addition to
reproducing the experimental data reasonably well-- the following two criteria
must be satisfied:

(i) one must obtain a consistent value of $V_{us}$ in the relevant
HSD, and

(ii) this latter value of $V_{us}$ must also be consistent with its
value of $K_{l3}$ decays, Eq.~(\ref{ec1}).

With the currently available experimental information the relevant HSD to
determine $V_{us}$ are $\Lambda \rightarrow p e \nu$, $\Sigma^{-} 
\rightarrow n e \nu$, and $\Xi^{-} \rightarrow \Lambda e \nu$\cite{part}.
This information in the form of decay rates, angular correlations, and spin
asymmetries is collected in Table~\ref{tabla1}. An alternative set of 
experimental data is constituted by the rates and the
measured $g_1/f_1$ ratios. However, this latter set is not as rich as the 
former and will not be used here.

In this paper we shall perform a detailed analysis of the success of the
predictions of references\cite{sch,krause,dono,ander} for HSD through the
values obtained for $V_{us}$, as explained above. In Sec.~II we shall 
briefly review the predictions of these references and we shall make the 
first determination of $V_{us}$. In Sec.~III we shall study the effect 
upon $V_{us}$ of the induced vector and axial-vector form factors $f_2$ 
and $g_2$, respectively. This study will give us a more precise 
determination of $V_{us}$. Sec.~IV will be reserved for discussions and 
conclusions. Our main result will be that only the predictions of 
Ref.\cite{ander} satisfy criteria (i) and (ii), in accordance  with the 
findings of a model independent analysis performed before\cite{garcia}.

\section{A FIRST DETERMINATION OF $V_{\lowercase{us}}$}

We shall refer to the calculations of references\cite{sch,krause,dono,ander}
as Models~I, II, III, and IV, respectively. Our interest in them arises 
from the fact that in each of them not only first order but second-order 
$SU(3)$ symmetry-breaking corrections to the leading vector form factor 
$f_1$ were calculated. In models I and III the corrections to the leading 
axial-vector form factor $g_1$ were also produced. The approaches and/or 
approximations used in going from one model to another are quite different.
In Model~I a relativistic quark model was used. Model~II made use of 
chiral perturbation and included corrections of $O(m_s)$. Model~III 
relied on the non-relativistic quark and bag models and included both 
wave-function mismatch and center of mass corrections. A similar approach 
treating solely center of mass corrections was analized in Ref.\cite{rat}.
Model~IV followed the lines of Model~II but it incorporated the more refined
corrections of $O(m_s^{3/2})$. The corresponding predictions for $f_1$ 
are reproduced in Table~\ref{tabla2}. They are displayed in the form of 
ratios $f_1/{f_1}^{SU(3)}$. The values of $g_1/{g_1}^{SU(3)}$ predicted 
by Models~I and III are displayed in Table~\ref{tabla3}. The symmetry 
limit values ${f_1}^{SU(3)}$ and ${g_1}^{SU(3)}$ correspond to the 
conserved vector current hypothesis (CVC) and the Cabibbo theory
predictions, respectively. A review of this last can be found in 
Ref.\cite{decay}.

For our analysis we shall include radiative corrections and the 
four-momentum transfer contributions of the form factors. The detailed 
expressions are given in Ref.\cite{decay}. None of the four models give 
the predictions for $f_2$ and $g_2$. In this section we shall assume for 
the several $f_2$ their CVC predictions and we shall keep each $g_2$ 
equal to zero, in accordance with the assumption of the absence of 
second-class currents. The other two induced form factors $f_3$ and $g_3$ 
can be safely ignored in the three decays we consider, because their 
contributions are proportional to the electron mass.

With this last information we can already make the first determination of
$V_{us}$ with Models~I and III. The values obtained are given in 
Tables~\ref{tabla4} and \ref{tabla5}, respectively. We show in these 
tables the values of $g_1$ used, but this time normalized with respect to 
$f_1$. The effects of considering only center of mass corrections in 
Model~III, as discussed in Ref. ~\cite{rat}, have been displayed in the 
entries within parentheses of Table~\ref{tabla5}. In the case of Models~II 
and IV we do not have the corresponding predictions for the $g_1$'s. We 
shall leave each one as a free parameter and then the results of 
Tables~\ref{tabla6} and \ref{tabla7} are obtained. In order to make a 
comparison on an equal footing of the four models we also leave the 
$g_1$'s as free parameters with Models~I and III. Tables~\ref{tabla8} 
and \ref{tabla9} are thus obtained.

Let us now look into the results obtained. The $f_1$ and $g_1$ form 
factors predicted by Models~I and III lead to values of $V_{us}$ that 
differ from one decay to another by more than three standard deviations, 
as can be seen in Tables~\ref{tabla4} and \ref{tabla5}. That is, 
criterion (i) above is not satisfied. In contrast, Models~II and IV do 
lead to values of $V_{us}$ in Tables~\ref{tabla6} and \ref{tabla7} that 
in each model are consistent with one another within a little bit more 
than one standard deviation. When the $g_1$'s are free parameters the new 
determinations of $V_{us}$ of Models~I and III given in 
Tables~\ref{tabla8} and \ref{tabla9} become consistent in each model, too.
The criterion (i) is satisfied by the four models when the $g_1$'s are 
allowed to be free parameters. The calculated $g_1$'s of Models~I and III 
seem to be ruled out by criterion (i). This is also confirmed by the high 
$\chi^2$ obtained when the $g_1$'s are fixed. However, in Model~III when 
only center of mass corrections are considered the $\chi^2$ of $\Lambda 
\to pe\nu$ is remarkably lowered although the value of $V_{us}$ obtained 
is increased with respect to the case when the wave-function mismatch 
corrections are included.

Concerning criterion (ii), we see that Models~I, II, and III give values of
$V_{us}$ that are systematically higher than the $K_{l3}$ value of
Eq.~(\ref{ec1}) close to three standard deviations in some cases or more than
three in other cases. In contrast, Model~IV gives systematically values 
of $V_{us}$ that are lower than Eq.~(\ref{ec1}). These values, however, 
are pretty close to Eq.~(\ref{ec1}).

The high $\chi^2$ in $\Lambda \to pe\nu$ in Tables~\ref{tabla4} and 
\ref{tabla5} is due mainly to $\alpha_{e\nu}$ and $\alpha_\nu$, whereas in 
Tables ~\ref{tabla6}--~\ref{tabla9} it comes mainly from $\alpha_e$ and 
$\alpha_\nu$. In the case of $\Sigma^-\to ne\nu$ the $\chi^2$ is also 
high and comes mainly from $\alpha_\nu$ and $\alpha_B$. Even when the 
$g_1$'s are used as free parameters, despite the appreciable lowering of 
$\chi^2$, a still rather high $\chi^2$ remains in $\Lambda \to pe \nu$ 
and $\Sigma^- \to ne \nu$.
 
Before drawing conclusions, it is important to consider the effect the induced
form factors $f_2$ and $g_2$ have upon the determination of $V_{us}$ and 
$\chi^2$. This we do in the next section.

\section{EFFECT OF THE INDUCED VECTOR AND AXIAL-VECTOR FORM FACTORS}

None of the four models under consideration here produced predictions for
$f_2$ and $g_2$. Nevertheless, it is necessary to study their relevance 
in determining $V_{us}$. We shall allow them to be free parameters, since 
inasmuch as they help reduce $\chi^2$ we may expect that experimental 
data, which certainly know of symmetry breaking corrections, will force 
them to move into the correct direction.

The CVC contributions of $f_2$ are already first-order symmetry breaking
contributions to the experimental observables of Table~\ref{tabla1}.
Accordingly, one should only consider first-order symmetry breaking of such
CVC predictions in order to take into account the second-order contributions
of the $f_2$. It is reasonable to allow the $f_2$'s to vary only up to 
20\% around the CVC values. This we shall do in steps, first by changing 
the $f_2$'s by $\pm 10\%$ and keeping them fixed while redoing the fits 
of the previous section and next by changing them by $\pm 20\%$ and 
repeating the whole procedure.

The results of this analysis are that practically no observable effects upon
the values of $V_{us}$ are seen to occur. Only the fourth digits are 
changed, without even affecting third digits by rounding up. There is no 
need to produce new tables with such negligible changes.

Due to the absence of second-class currents, the $g_2$ are all zero in the
symmetry limit. They will be rendered non-zero by $SU(3)$ symmetry breaking.
As in the case of the $f_2$, the first-order corrections to them will amount
to second-order contributions to the observables. We shall introduce 
fixed values of the $g_2$'s first of $\pm 0.10$ and next by $\pm 0.20$ 
and redo all the fits of Sec.~II. These changes seem to be of reasonable 
size according to the estimations of Refs.\cite{carson} and \cite{rath}.
The $g_2$'s do lead observable changes.

Models~I and III with $f_1$ and $g_1$ fixed at their predictions give values
of $V_{us}$ in $\Lambda \rightarrow p e \nu$ that come closer to
Eq.~(\ref{ec1}), but still with high $\chi^2$ --meaning that the corresponding
experimental data are not satisfactorily reproduced. Also the dispersion 
of the values of $V_{us}$ from the three decays, although somewhat 
mitigated is not corrected either. All this is collected in 
Tables~\ref{tabla10} and \ref{tabla11}. The effect of dropping the 
wave-function mismatch corrections of Model~III is displayed in the 
entries within parentheses of Table~\ref{tabla11}. Again an appreciable 
lowering of $\chi^2$ is seen in $\Lambda \to pe\nu$ and also in 
$\Sigma^- \to ne\nu$, but at the expense of increasing $V_{us}$ with 
respect to the corresponding values of $V_{us}$ when such corrections 
are included.

When the $g_1$'s are allowed to vary then Models~I and III improve their
agreement with experiment, the corresponding $\chi^2$'s are noticeably 
reduced. This can be seen in Tables~\ref{tabla12} and \ref{tabla13}.
But the values of $V_{us}$ are increased to the extent that none is any 
longer compatible with Eq.~(\ref{ec1}). This situation repeats itself for 
Model~II in Table~\ref{tabla14}. In contrast, the values of $V_{us}$ 
obtained with Model~IV are fairly stable with respect to changes of 
$g_2$. Actually, as seen in Table~\ref{tabla15} they tend to increase 
with respect to the corresponding entries of Table~\ref{tabla7}, which 
is in the right direction towards Eq.~(\ref{ec1}).

Concerning the agreement with experiment we observe that a further 
lowering to an acceptable value of $\chi^2$ is obtained in $\Sigma^- \to 
ne\nu$ as an effect of a non-zero $g_2$. However, this lowering is not 
observed in the $\chi^2$ of $\Lambda \to pe\nu$, which remains at around 
10 through Tables~\ref{tabla12} -- ~\ref{tabla15}. This effect may be 
due to some experimental inconsistency of the value of $\alpha_\nu$, 
which contributes 7 to $\chi^2$, with the other asymmetries. If 
this $\alpha_\nu$ is left out the same $V_{us}$ is obtained along with 
practically the same error bars. For example, with the $f_1$ of Model~III 
and with variable $g_1/f_1$, one obtains $V_{us} = 0.2220 \pm 0.0035, 
0.2261 \pm 0.0035$, and $0.2302 \pm 0.0035$ for $\Delta g_2 = -0.20, 0.0$, 
and $+0.20$, respectively. The corresponding $\chi^2$'s are 4.30, 4.1, and 
4.0, which represents a considerable reduction with respect to the 
corresponding $\chi^2$'s in Table~\ref{tabla13}; these new $\chi^2$'s 
indicate a very good agreement with other four observables in $\Lambda 
\to p e \nu$. The same pattern repeats itself when $\alpha_\nu$ is left 
out in the comparison of the other models. In view of this situation we 
shall keep the several tables as they are. The high $\chi^2$ of $\Lambda 
\to p e\nu$ should serve as a remainder that some problem exists in this 
decay. It is not idle to insist that new measurements in this decay 
should be most welcome.  

The combined effect of simultaneous changes of $f_2$ and $g_2$ leads to the
same results of Tables~\ref{tabla10} -- \ref{tabla15}, except for minor 
changes in the fourth digits of the several values of $V_{us}$. Again 
there is no need to produce tables to show this. Let us pass to the last 
section.

\section{DISCUSSION AND CONCLUSIONS}

Throughout our study, we notice that the values obtained for $V_{us}$ 
are very model dependent. We also notice that, except for one model, the 
values of $V_{us}$ are inconsistent with each other within the same 
model. These observations render inadmissible to quote a consistent 
average value from HSD.

However, since the dispersion of the values of $V_{us}$ in each of the 
three decays is mitigated in all the models when one allows the $g_1$ to 
be free parameters, one may quote an average value of the $V_{us}$ 
obtained with each model by selecting the appropriate sign of $\Delta 
g_2$ that lowers most the corresponding $\chi^2$. That is, we accept 
that criterion (i) is more or less satisfied by each model. These 
averages are collected in Table~\ref{tabla16}. We have also included 
there the averages of the case $\Delta g_2 = 0$. This last table allows 
us to better appreciate  how criterion (ii) is satisfied or not.

Looking at the averages obtained for $V_{us}$ with each model, one readily
sees that Models~I, II, and III are far from satisfying criterion (ii), while
Model~IV satisfies it remarkably well. From this point of view, it 
becomes very clear that the criteria discussed in the introduction indeed 
serve as quite stringent discriminating tools between different models 
and/or approximations. Our main conclusion in this regard is that of the 
four models that provide second-order symmetry breaking corrections to 
the $f_1$'s only Model~IV of Ref.\cite{ander} is acceptable.

This conclusion allows us to quote the best value of $V_{us}$ that can be
obtained from Model~IV, namely,

\begin{equation}
V_{us} = 0.2176 \pm 0.0026 \, .
\label{ec2}
\end{equation}

\noindent
Since this value is statistically in very good agreement with the $K_{l3}$
one of Eq.~(\ref{ec1}), we can average both and get

\begin{equation}
V_{us}^{\rm AV} = 0.2185 \pm 0.0014 \, .
\label{ec3}
\end{equation}

The determination of $V_{us}$ in Eq.~(\ref{ec2}) is quite acceptable in the
light of the model independent analysis of Ref.\cite{garcia}. Although we 
have committed ourselves with the predictions of Model~IV for the 
$f_1$'s, the rest of the form factors was dealt with in a model-independent
fashion.

This last remark brings us to our closing comments. One cannot yet 
consider the theoretical issues as closed. It is most important that 
within the same Model~IV used to calculate the $f_1$'s the other 
relevant form factors be also computed. The values displayed for these 
form factors in Tables~\ref{tabla7} and \ref{tabla15} may provide useful 
guidance for this enterprise. Our analysis of Sec.~III shows that 
detailed values of the $f_2$'s are not relevant and thus these HSD do 
not provide useful guidance for their calculation. It should be found 
elsewhere. Also, as pointed out in Ref.~\cite{rat96} a viable model of 
$SU(3)$ breaking should be able to predict the $\Delta S = 0$ modes, 
$\Sigma ^\pm \to \Lambda e \nu$. Only if the predictions for $\Delta S = 
0$ and $\Delta S \neq 0$ decays are simultaneously correct should one 
consider Model~IV completely successful. 

\acknowledgments

The authors wish to express their gratitude to CONACyT for partial
support.

\begin{table}
\caption{
Experimental data for the three relevant HSD. The units of $R$ are $10^6$ 
${\rm s^{-1}}$. }
\label{tabla1}
\begin{tabular}{
c
r@{.}l@{\,$\pm$\,}r@{.}l
r@{.}l@{\,$\pm$\,}r@{.}l
r@{.}l@{\,$\pm$\,}r@{.}l
}
&
\multicolumn{4}{c}{$\Lambda \rightarrow p e \nu$} &
\multicolumn{4}{c}{$\Sigma^{-} \rightarrow n e \nu$} &
\multicolumn{4}{c}{$\Xi^{-} \rightarrow \Lambda e \nu$}
\\
\tableline
$R$ &
3 & 169 & 0 & 058 &
6 & 876 & 0 & 235 &
3 & 36 & 0 & 19
\\
$\alpha_{e\nu}$ &
$-$0 & 019 & 0 & 013 &
0 & 347 & 0 & 024 &
0 & 53 & 0 & 10
\\
$\alpha_e$ &
0 & 125 & 0 & 066 &
$-$0 & 519 & 0 & 104 &
\multicolumn{4}{c}{}
\\
$\alpha_\nu$ &
0 & 821 & 0 & 060 &
$-$0 & 230 & 0 & 061 &
\multicolumn{4}{c}{}
\\
$\alpha_B$ &
$-$0 & 508 & 0 & 065 &
0 & 509 & 0 & 102 &
\multicolumn{4}{c}{}
\\
$A$ &
\multicolumn{4}{c}{} &
\multicolumn{4}{c}{} &
0 & 62 &  0 & 10
\\
$g_1/f_1$ &
0 & 718 & 0 & 015 &
$-$0 & 340 & 0 & 017 &
0 & 25 & 0 & 05
\\
\end{tabular}
\end{table}

\begin{table}
\caption{
$SU(3)$ breaking for $f_1$.
The values correspond to the ratio $f_1/{f_1}^{\rm SU(3)}$.
}
\label{tabla2}
\begin{tabular}{cdddd}
Decay & Model I & Model II & Model III & Model IV \\
\tableline
$\Lambda\rightarrow p e \nu$ & 0.976 & 0.943 & 0.987 & 1.024  \\
$\Sigma^{-}\rightarrow n e \nu$ & 0.975 & 0.987 & 0.987 & 1.100  \\
$\Xi^{-}\rightarrow \Lambda e \nu$ & 0.976 & 0.957 & 0.987 & 1.059  \\
\end{tabular}
\end{table}

\begin{table}
\caption{
SU(3) breaking for $g_1$.
The values correspond to the ratio $g_1/{g_1}^{\rm SU(3)}$.
In parentheses, the breaking pattern of Model~III including only
center of mass corrections is given.
}
\label{tabla3}
\begin{tabular}{cdc}
Decay & Model I & Model III \\
\tableline
$\Lambda\rightarrow pe\nu$ & 1.072 & 1.050\ \ (0.9720) \\
$\Sigma^{-}\rightarrow ne\nu$ & 1.056 & 1.040\ \ (0.9628) \\
$\Xi^{-}\rightarrow \Lambda e\nu$ & 1.072 & 1.003\ \ (0.9287) \\
\end{tabular}
\end{table}

\begin{table}
\caption{
Values of $V_{us}$ within the SB proposed by Model~I.
Both breaking patterns for $f_1$ and $g_1$ were used.
}
\label{tabla4}
\begin{tabular}{ccdd}
Decay & $V_{us}$ & $g_1/f_1$ & $\chi^2$ \\
\tableline
$\Lambda\rightarrow pe\nu$ & 0.2133 $\pm$ 0.0020 & 0.8019 & 38.54 \\
$\Sigma^{-}\rightarrow ne\nu$ & 0.2318 $\pm$ 0.0040 & $-$0.3529 & 8.95 \\
$\Xi^{-}\rightarrow\Lambda e\nu$ & 0.2434 $\pm$ 0.0068 & 0.2221 & 1.40 \\
\end{tabular}
\end{table}

\begin{table}
\caption{
Values of $V_{us}$ within the SB proposed by Model~III.
Both breaking patterns for $f_1$ and $g_1$ were used.
In parentheses, below each entry, the corresponding values of $V_{us}$, 
$g_1/f_1$, and $\chi^2$ considering only center of mass corrections are 
given.
}
\label{tabla5}
\begin{tabular}{ccdd}
Decay & $V_{us}$ & $g_1/f_1$ & $\chi^2$ \\
\tableline
$\Lambda\rightarrow pe\nu$ & $0.2153 \pm 0.0020$ & 0.7767 & 25.43 \\
& $(0.2258 \pm 0.0021)$ & (0.7190) & (10.85) \\
$\Sigma^{-}\rightarrow ne\nu$ & $0.2307 \pm 0.0040$ & $-$0.3433 & 7.92 \\
& $(0.2351 \pm 0.0041)$ & ($-$0.3178) & (8.89) \\
$\Xi^{-}\rightarrow \Lambda e\nu$ & $0.2429 \pm 0.0068$ & 0.2055 & 2.42 \\
& $(0.2449 \pm 0.0069)$ & (0.1903) & (3.62) \\
\end{tabular}
\end{table}

\begin{table}
\caption{
Values of $V_{us}$ within the SB proposed by Model~II, with $g_1$ as free
parameter.
}
\label{tabla6}
\begin{tabular}{ccddd}
Decay & $V_{us}$ & $g_1$ & $g_1/f_1$ & $\chi^2$ \\
\tableline
$\Lambda\rightarrow pe\nu$ &
0.2372 $\pm$ 0.0037 &
$-$0.8250 &
0.7142 &
10.79 \\
$\Sigma^{-}\rightarrow ne\nu$ &
0.2320 $\pm$ 0.0049 &
0.3312 &
$-$0.3356 &
7.70 \\
$\Xi^{-}\rightarrow\Lambda e\nu$ &
0.2396 $\pm$ 0.0108 &
0.3264 &
0.2784 &
$6\times 10^{-3}$ \\
\end{tabular}
\end{table}

\begin{table}
\caption{
Values of $V_{us}$ within the SB proposed by Model~IV, with $g_1$ as free
parameter.
}
\label{tabla7}
\begin{tabular}{ccddd}
Decay & $V_{us}$ & $g_1$ & $g_1/f_1$ & $\chi^2$ \\
\tableline
$\Lambda\rightarrow pe\nu$ & 0.2183 $\pm$ 0.0034 & $-$0.8974 & 0.7155 & 10.77
\\
$\Sigma^{-}\rightarrow ne\nu$ & 0.2082 $\pm$ 0.0044 & 0.3694 & $-$0.3358 &
6.73 \\
$\Xi^{-}\rightarrow\Lambda e\nu$ & 0.2165 $\pm$ 0.0098 & 0.3611 & 0.2784 &
$6\times 10^{-3}$ \\
\end{tabular}
\end{table}

\begin{table}
\caption{
Values of $V_{us}$ within the SB proposed by Model~I, with $g_1$ as free
parameter.
}
\label{tabla8}
\begin{tabular}{ccddd}
Decay & $V_{us}$ & $g_1$ & $g_1/f_1$ & $\chi^2$ \\
\tableline
$\Lambda\rightarrow pe\nu$ & 0.2291 $\pm$ 0.0036 & $-$0.8545 & 0.7148 & 10.78
\\
$\Sigma^{-}\rightarrow ne\nu$ & 0.2349 $\pm$ 0.0049 & 0.3271 & $-$0.3355 &
7.82 \\
$\Xi^{-}\rightarrow\Lambda e\nu$ & 0.2349 $\pm$ 0.0106 & 0.3328 & 0.2784 &
$6\times 10^{-3}$ \\
\end{tabular}
\end{table}

\begin{table}
\caption{
Values of $V_{us}$ within the SB proposed by Model~III, with $g_1$ as free
parameter.
}
\label{tabla9}
\begin{tabular}{ccddd}
Decay & $V_{us}$ & $g_1$ & $g_1/f_1$ & $\chi^2$ \\
\tableline
$\Lambda\rightarrow pe\nu$ & 0.2265 $\pm$ 0.0035 & $-$0.8643 & 0.7149 & 10.78
\\
$\Sigma^{-}\rightarrow ne\nu$ & 0.2320 $\pm$ 0.0049 & 0.3312 & $-$0.3356 &
7.70 \\
$\Xi^{-}\rightarrow\Lambda e\nu$ & 0.2323 $\pm$ 0.0105 & 0.3366 & 0.2784 &
$6\times 10^{-3}$ \\
\end{tabular}
\end{table}

\begin{table}
\squeezetable
\caption{
Values of $V_{us}$ within the SB proposed by Model~I. $f_1$ and
$g_1$ are fixed. $g_2$ are non-zero.
}
\label{tabla10}
\begin{tabular}{ccdcdcdcd}
$\Delta g_2$ &
\multicolumn{2}{c}{$-0.20$} &
\multicolumn{2}{c}{$-0.10$} &
\multicolumn{2}{c}{$+0.10$} &
\multicolumn{2}{c}{$+0.20$} \\
\tableline
Decay &
$V_{us}$ &
$\chi^2$ &
$V_{us}$ &
$\chi^2$ &
$V_{us}$ &
$\chi^2$ &
$V_{us}$ &
$\chi^2$ \\
\tableline
$\Lambda\rightarrow pe\nu$ & $0.2163 \pm 0.0020$ & 19.7 &
$0.2148 \pm 0.0020$ & 27.7 & $0.2118 \pm 0.0019$ & 52.1 &
$0.2103 \pm 0.0019$ & 68.1 \\
$\Sigma^{-}\rightarrow ne\nu$ & $0.2266 \pm 0.0039$ & 19.9 &
$0.2292 \pm 0.0040$ & 12.2 & $0.2343 \pm 0.0040$ &  9.9 &
$0.2368 \pm 0.0041$ & 15.0 \\
$\Xi^{-}\rightarrow\Lambda e\nu$ & $0.2409 \pm 0.0068$ & 0.8 &
$0.2421 \pm 0.0068$ & 1.0 & $0.2445 \pm 0.0069$ & 1.8 & $0.2457 \pm 0.0069$ &
2.3 \\
\end{tabular}
\end{table}

\begin{table}
\squeezetable
\caption{
Values of $V_{us}$ within the SB proposed by Model~III. $f_1$ and $g_1$ 
are fixed. $g_2$ are non-zero. In parentheses, below each entry, the 
corresponding values of $V_{us}$ and $\chi^2$ considering only center of 
mass corrections are given. }
\label{tabla11}
\begin{tabular}{ccdcdcdcd}
$\Delta g_2$ &
\multicolumn{2}{c}{$-0.20$} &
\multicolumn{2}{c}{$-0.10$} &
\multicolumn{2}{c}{$+0.10$} &
\multicolumn{2}{c}{$+0.20$} \\
\tableline
Decay &
$V_{us}$ &
$\chi^2$ &
$V_{us}$ &
$\chi^2$ &
$V_{us}$ &
$\chi^2$ &
$V_{us}$ &
$\chi^2$ \\
\tableline
$\Lambda \rightarrow p e \nu$ &
$0.2183 \pm 0.0020$ & 13.4 & $0.2168 \pm 0.0020$ & 18.0 &
$0.2138 \pm 0.0020$ & 35.7 & $0.2123 \pm 0.0020$ & 48.7 \\
& $(0.2290 \pm 0.0021)$ & (17.0) & $(0.2274 \pm 0.0021)$ & (12.2) &
$(0.2241 \pm 0.0021)$ & (12.8) & $(0.2225 \pm 0.0020)$ & (17.9) \\
$\Sigma^{-} \rightarrow n e \nu$ &
$0.2256 \pm 0.0039$ & 15.2 & $0.2282 \pm 0.0039$ &  9.5 &
$0.2331 \pm 0.0040$ & 10.5 & $0.2355 \pm 0.0041$ & 16.9 \\
& $(0.2301 \pm 0.0040)$ & (7.2) & $(0.2327 \pm 0.0040)$ & (6.0) &
$(0.2375 \pm 0.0041)$ & (15.6) & $(0.2398 \pm 0.0041)$ & (25.9) \\
$\Xi^{-} \rightarrow \Lambda e \nu$ &
$0.2406 \pm 0.0068$ &  1.5 & $0.2418 \pm 0.0068$ & 1.9 &
$0.2440 \pm 0.0069$ &  3.0 & $0.2450 \pm 0.0069$ &  3.6 \\
& $(0.2427 \pm 0.0068)$ & (2.5) & $(0.2438 \pm 0.0069)$ & (3.0) &
$(0.2459 \pm 0.0069)$ & (4.3) & $(0.2469 \pm 0.0069)$ & (5.0) \\
\end{tabular}
\end{table}

\begin{table}
\squeezetable
\caption{
Values of $V_{us}$ within the SB proposed by Model~I. The $g_1$ are free 
and $g_2$ are non-zero. In parentheses, below the entries for $V_{us}$, 
the corresponding $g_1$ are also given.
}
\label{tabla12}
\begin{tabular}{c cd  cd  cd  cd}
$\Delta g_2$ &
\multicolumn{2}{c}{$-0.20$} &
\multicolumn{2}{c}{$-0.10$} &
\multicolumn{2}{c}{$+0.10$} &
\multicolumn{2}{c}{$+0.20$} \\
\tableline
Decay &
$V_{us}$ &
$\chi^2$ &
$V_{us}$  &
$\chi^2$ &
$V_{us}$ &
$\chi^2$ &
$V_{us}$ &
$\chi^2$ \\
\tableline
$\Lambda \rightarrow p e \nu$ &
$0.2248 \pm 0.0036$ & 11.6 & $0.2270 \pm 0.0036$ & 11.2 &
$0.2312 \pm 0.0036$ & 10.4 & $0.2333 \pm 0.0036$ & 10.0 \\
& $(-0.9025)$ & & $(-0.8784)$ & & $(-0.8308)$ & & $(-0.8075)$ & \\
$\Sigma^{-} \rightarrow n e \nu$ &
$0.2377 \pm 0.0049$ & 4.4 & $0.2364 \pm 0.0049$ & 6.0 &
$0.2333 \pm 0.0050$ & 9.8 & $0.2316 \pm 0.0050$ & 12.0 \\
& $(0.2835)$ & & $(0.3051)$ & & $(0.3496)$ &  & $(0.3726)$ & \\
$\Xi^{-} \rightarrow \Lambda e \nu$ &
$0.2349 \pm 0.0104$ & 0.0 & $0.2349 \pm 0.0105$ & 0.0 &
$0.2349 \pm 0.0108$ & 0.0 & $0.2349 \pm 0.0109$ & 0.0 \\
& $(0.3123)$ & & $(0.3226)$ & & $(0.3431)$ & & $(0.3534)$ & \\
\end{tabular}
\end{table}

\begin{table}
\squeezetable
\caption{
Values of $V_{us}$ within the SB proposed by Model~III. The $g_1$ are 
free and the $g_2$ are non-zero. In parentheses, below the entries for 
$V_{us}$, the corresponding $g_1$ are also given.
}
\label{tabla13}
\begin{tabular}{c cd  cd  cd  cd}
$\Delta g_2$ &
\multicolumn{2}{c}{$-0.20$} &
\multicolumn{2}{c}{$-0.10$} &
\multicolumn{2}{c}{$+0.10$} &
\multicolumn{2}{c}{$+0.20$} \\
\tableline
Decay &
$V_{us}$ &
$\chi^2$ &
$V_{us}$ &
$\chi^2$ &
$V_{us}$ &
$\chi^2$ &
$V_{us}$ &
$\chi^2$ \\
\tableline
$\Lambda \rightarrow p e \nu$ &
$0.2223 \pm 0.0035$ &
11.6 &
$0.2244 \pm 0.0036$ &
11.2 &
$0.2286 \pm 0.0035$ &
10.4 &
$0.2307 \pm 0.0035$ &
10.0 \\
& $(-0.9123)$ &
& $(-0.8882)$ &
& $(-0.8407)$ &
& $(-0.8173)$ & \\
$\Sigma^{-} \rightarrow n e \nu$ &
$0.2348 \pm 0.0048$ &
4.4 & $0.2335 \pm 0.0049$ &
5.9 & $0.2305 \pm 0.0049$ &
9.7 & $0.2288 \pm 0.0049$ &
11.8 \\
& $(0.2876)$ &
& $(0.3092)$ &
& $(0.3537)$ &
& $(0.3767)$ & \\
$\Xi^{-} \rightarrow \Lambda e \nu$ &
$0.2323 \pm 0.0103$ &
0.0 & $0.2323 \pm 0.0104$ &
0.0 & $0.2322 \pm 0.0106$ &
0.0 & $0.2322 \pm 0.0108$ &
0.0 \\
& $(0.3161)$ &
& $(0.3263)$ &
& $(0.3468)$ &
& $(0.3571)$ & \\
\end{tabular}
\end{table}

\begin{table}
\squeezetable
\caption{
Values of $V_{us}$ within the SB proposed by Model~II.
The $g_1$ are free and the $g_2$ are non-zero
In parentheses, below the entries for $V_{us}$, the corresponding $g_1$ are
also given.
}
\label{tabla14}
\begin{tabular}{c cd  cd  cd  cd}
$\Delta g_2$ &
\multicolumn{2}{c}{$-0.20$} &
\multicolumn{2}{c}{$-0.10$} &
\multicolumn{2}{c}{$+0.10$} &
\multicolumn{2}{c}{$+0.20$} \\
\tableline
Decay &
$V_{us}$ &
$\chi^2$ &
$V_{us}$ &
$\chi^2$ &
$V_{us}$ &
$\chi^2$ &
$V_{us}$ &
$\chi^2$ \\
\tableline
$\Lambda \rightarrow p e \nu$ &
$0.2325 \pm 0.0037$ &
11.6 &
$0.2349 \pm 0.0037$ &
11.2 &
$0.2394 \pm 0.0037$ &
10.4 &
$0.2417 \pm 0.0037$ &
9.9 \\
& $(-0.8730)$ &
& $(-0.8489)$ &
& $(-0.8013)$ &
& $(-0.7780)$ & \\
$\Sigma^{-} \rightarrow n e \nu$ &
$0.2348 \pm 0.0048$ &
4.4 &
$0.2335 \pm 0.0049$ &
5.9 &
$0.2305 \pm 0.0049$ &
9.7 &
$0.2288 \pm 0.0049$ &
11.8 \\
& $(0.2876)$ &
& $(0.3092)$ &
& $(0.3537)$ &
& $(0.3767)$ & \\
$\Xi^{-} \rightarrow \Lambda e \nu$ &
$0.2396 \pm 0.0106$ &
0.0 &
$0.2396 \pm 0.0107$ &
0.0 &
$0.2395 \pm 0.0110$ &
0.0 &
$0.2395 \pm 0.0111$ &
0.0 \\
& $(0.3059)$ &
& $(0.3162)$ &
& $(0.3366)$ &
& $(0.3469)$ & \\
\end{tabular}
\end{table}

\begin{table}
\squeezetable
\caption{
Values of $V_{us}$ within the SB proposed by Model~IV.
The $g_1$ are free and the $g_2$ are non-zero
In parentheses, below the entries for $V_{us}$, the corresponding $g_1$ is
also given.
}
\label{tabla15}
\begin{tabular}{c cd  cd  cd  cd}
$\Delta g_2$ &
\multicolumn{2}{c}{$-0.20$} &
\multicolumn{2}{c}{$-0.10$} &
\multicolumn{2}{c}{$+0.10$} &
\multicolumn{2}{c}{$+0.20$} \\
\tableline
Decay &
$V_{us}$ &
$\chi^2$ &
$V_{us}$ &
$\chi^2$ &
$V_{us}$ &
$\chi^2$ &
$V_{us}$ &
$\chi^2$
\\
\tableline
$\Lambda \rightarrow p e \nu$ &
$0.2144 \pm 0.0034$ &
11.5 &
$0.2164 \pm 0.0034$ &
11.2 &
$0.2200 \pm 0.0037$ &
10.4 &
$0.2222 \pm 0.0034$ &
10.0 \\
& $(-0.9454)$ &
& $(-0.9212)$ &
& $(-0.8748)$ &
& $(-0.8503)$ & \\
$\Sigma^{-} \rightarrow n e \nu$ &
$0.2104 \pm 0.0043$ &
 3.9 &
$0.2093 \pm 0.0044$ &
 5.2 &
$0.2070 \pm 0.0044$ &
 8.4 &
$0.2056 \pm 0.0044$ &
10.2 \\
& $(0.3258)$ &
& $(0.3474)$ &
& $(0.3919)$ &
& $(0.4147)$ & \\
$\Xi^{-} \rightarrow \Lambda e \nu$ &
$0.2165 \pm 0.0096$ &
0.0 &
$0.2165 \pm 0.0097$ &
0.0 &
$0.2165 \pm 0.0099$ &
0.0 &
$0.2164 \pm 0.0100$ &
0.0 \\
& $(0.3406)$ &
& $(0.3508)$ &
& $(0.3714)$ &
& $(0.3816)$ & \\
\end{tabular}
\end{table}

\begin{table}
\caption{
Values of $V_{us}$ obtained within different $SU(3)$ SB models with 
changes in $g_2$. The rates and angular coefficients were used.
}
\label{tabla16}
\begin{tabular}{dcccc}
$\Delta g_2$ &
Model I &
Model II &
Model III &
Model IV
\\
\tableline
$=$ 0 &
$0.2314 \pm 0.0028$ &
$0.2356 \pm 0.0028$ &
$0.2286 \pm 0.0027$ &
$0.2147 \pm 0.0026$ \\
$\neq$ 0 &
$0.2348 \pm 0.0028$ &
$0.2392 \pm 0.0028$ &
$0.2321 \pm 0.0027$ &
$0.2176 \pm 0.0026$ \\
\end{tabular}
\end{table}

\end{document}